\documentclass{aa}
\usepackage{natbib}
\bibpunct{(}{)}{;}{a}{}{,}
\usepackage{amssymb}
\usepackage{graphicx}
\usepackage{txfonts}
\begin{document}
%
%
   \title{$p$-mode frequencies in solar-like stars:
\thanks{Based on observations obtained at
 the Observatoire de Haute-Provence (CNRS, France) and at the Whipple Observatory (Arizona, USA)}
}

    \subtitle{I. Procyon~A}

   \author{M.~Marti\'{c}\inst{1}, J.-C.~Lebrun\inst{1}, T.~Appourchaux\inst{2}
         \and
         S. G.~Korzennik\inst{3}
          }

   \offprints{M. Marti\'{c}}
   \mail{milena@aerov.jussieu.fr}

   \institute{Service d'A\'eronomie du CNRS, BP No 3, 91371
    Verri\`eres le Buisson, France
        \and
            Science Payload and Advanced Concept Office of ESA, ESTEC, AG Noordwijk, NL-2200 
         \and
            Harvard-Smithsonian Center for Astrophysics, 60 Garden Street, Cambridge, MA-02138, USA
  }

   \date{Received October 24, 2003; accepted January 27, 2004}

   \abstract{
As a part of an on-going program to explore the signature of $p$-modes 
in solar-like stars by means of high-resolution absorption line spectroscopy,
 we have studied four stars ($\alpha$\,CMi, $\eta$\,Cas A, $\zeta$\,Her A and $\beta$\,Vir).
 We present here new results from two-site observations of Procyon A acquired over twelve nights in 1999. Oscillation frequencies for $l$=1 and 0 (or 2) $p$-modes are detected in the power spectra of these Doppler shift measurements. A frequency analysis points out the difficulties of the  classical asymptotic theory in representing the $p$-mode spectrum of Procyon A. 
               }

\titlerunning{$p$-mode frequencies on Procyon A}
\authorrunning{Marti\'c et al.}
   \maketitle
%

\section{Introduction}

Precise measurements of the frequencies and amplitudes of the solar five-minute oscillations have
 led to detailed inferences about the solar internal structure \citep[cf. review paper by][]{Christ02}.
 Because of the wealth of knowledge obtained from helioseismology there is a strong impetus
  to use the pulsation analysis techniques to probe the interiors of solar-like stars.
 From the ground, the application of seismic techniques to these stars is difficult because
  of extremely small variations in intensity and velocity associated with $p$-modes
   and the  need to have an adequate temporal coverage to resolve the modes.
 To detect such small amplitudes, photometric methods
\citep{Gilliland93} suffer from scintillation noise unless conducted from space, e.g. MOST \citep{Walker03,Matthews02} or the future COROT \citep{Baglin03}
 and Eddington \citep{RoxburghFavata03} missions. 
Recent developments in high-precision spectrometric techniques and, in particular, Doppler spectroscopy 
have led to the detection of $p$-mode frequencies in individual solar-like stars \citep[e.g.,][]{Martic99,Bedding01,Martic01b,BouchyCarrier02,Frandsen02,Kjeldsen03,CarrierBourban03}. A recent review of these and other
measurements has been given by \citet{BeddingKjeldsen03}.

One of the most interesting target for solar-like seismology is Procyon A
 ($\alpha$\,CMi, HR 2943, HD61421), a F5\,IV star with $m_{\rm v}$=0.363 at a distance of only 3.53\,pc. Its fundamental stellar parameters are now well determined. Procyon A is in a 40-year
 period visual binary system, sharing the systemic velocity with a white dwarf. 
Procyon's astrometric orbit was recently updated by \citet{Girard00}. With a parallax, $\Pi = 283.2 \pm 1.5$ mas, they estimated a mass of $1.470\pm0.045 M_{\odot}$, which agrees well with previous excellent predictions from stellar interior models
 \citep{GuentherDemarque93}.
 In addition, \citet{Mozurkewich91}, using optical interferometry, measured a stellar angular diameter $\theta$ = 5.51$\pm 0.5$\,mas.
 Adopting the very precisely measured parallax
by \textit{Hipparcos},
 \mbox{$\Pi = 285.93 \pm\,0.88$\,mas}, \citet{Prieto02} derived 
a slightly lower mass of 1.42\,$\pm\,0.06 \rm\thinspace M_{\odot}$, 
a radius  R/$\rm\thinspace R_{\odot} = 2.071\,\pm\,0.02$ and a gravity log\,$g=3.96\,\pm0.02$\,(in cgs units).

An exploratory single-site Doppler observation of Procyon A by \citet{Martic99}, hereafter Paper I, revealed
 the presence of several frequencies that fell in a comb-like pattern with roughly equal spacing of $55\,\mu$Hz
 in the region of excess power around 1\,mHz
  in the power spectrum. 
Because of the large daytime gaps in these single-site measurements,
  it was not possible to identify these frequencies unambiguously. Note however that the set of  frequencies
   reported in Paper I were determined very close ($\leq 2\,\mu$Hz)
    to the model frequencies published independently
    by \citet{Chaboyer99}.
  
  Following the findings of these exploratory studies we have carried out a two-site campaign of Procyon A 
  to identify as many 
  $p$-modes as possible without alias ambiguities. 
In section~\ref{section_obs}, we describe the context of the observing runs, in section~\ref{section_period} we present the periodograms
 of the time sequences from 
coordinated two-site measurements.
In  sections~\ref{section_echdiag} and \ref{section_modes}, we outline the different techniques we used to search for the $p$-modes in the power spectra. 
In the last sections, we present the principal mode characteristics (large and small separations, amplitudes)
  and compare the results with the existing models.

\section{Observations and Data reduction}
   \begin{figure}
\vspace{0cm}
\resizebox{\hsize}{!}{\includegraphics{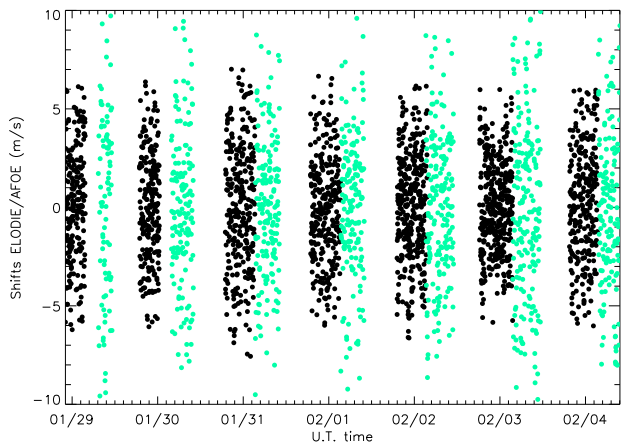}}
\vspace{0cm}
      \caption[]{Doppler shift measurements over seven nights from two-sites observations of Procyon A (black dots: ELODIE data, green dots: AFOE data).}
         \label{Fig_obs}
   \end{figure}

\label{section_obs}

The observations were carried out with the ELODIE spectrograph mounted on the 1.93m OHP telescope, France
and the Advanced Fibre-Optic Echelle (AFOE) spectrometer 
 located at the 1.5m (60\arcsec) Tillinghast telescope of the Whipple Observatory, Arizona, USA.
 The ELODIE and AFOE are fiber-fed cross-dispersed echelle spectrographs
  \citep{Baranne96,Brown94} that were optimized
  for precise and stable Doppler measurements. Both systems use double-fiber scramblers. 
  
  As configured for these coordinated measurements, AFOE
 covered 160\,nm of spectrum lying between 393 and 665\,nm (23 echelle orders) at a resolution of roughly 
 $R\sim32000$. The wavelength reference used for these observations was a ThAr emission-line spectrum
 brought into the spectrograph by a second optical fiber. The reduction steps for AFOE data
 are explained in \citet{Brown97}. A total number of 1127 extracted spectra,
 with sampling rate of about 140\,s, has been used for this Doppler velocity time sequence analysis.

  The ELODIE data were obtained as a continuous series of raw CCD (Tk 1024)
   frames with a sampling rate of about 100\,s. We built an
   asteroseismic mode to automatically acquire long uninterrupted sequences of simultaneous
    stellar and reference (Fabry-Perot) exposures. 
The advantage of using the closely-spaced channelled spectrum from a fixed (ZERODUR) Fabry-Perot
      interferometer is that it gives the best possible reference even for very short exposure times
(30\,s for Procyon)
       and allows us to monitor with a high precision the spectrograph instabilities. In our configuration,
 the reduction process produces for each  
CCD frame a set of stellar and Fabry-Perot spectra, at a resolution of $R\sim42000$,
 interleaved into 67 echelle orders, each covering 5.25\,nm or in total a wavelength
 domain from 390.6 to 681.1\,nm. 
  
 The radial velocity computation algorithm is based on the method
  proposed by \citet{Connes85} for the computation of the very small residual shifts with an absolute accelerometer.
   Our algorithm was first tested on $\psi$\,UMa,  HR4335 \citep[see][]{Connes96}
   and optimized with a spline interpolation for the iterative Doppler shift calculation
    to take account of the larger shifts induced by earth motion
     and instrument drifts during the night (cf. Paper I).

 The formalism of the Connes' algorithm was summarized in \citet{Bouchy01}. 
It was recently used as the "optimum weight procedure" by
 \citet{BouchyCarrier02}
  for the detection of the oscillations of $\alpha$\,Cen and by \citet{Frandsen02} for the giant star $\xi$\,Hya. 
   We should however note that the use of this method, for close echelle orders, 
is very sensitive to the star flux variation in the course of the night
    that induces an intensity-velocity correlation and in consequence a larger amplitude of the oscillation signal.

\begin{table}
\begin{tabular}{|llll|llll|}
            \hline
            \noalign{\smallskip}
       \multicolumn{4}{c}{ELODIE (1999)} & \multicolumn{4}{c}{AFOE (1999)}   \\
            Date   &  Nbr  & T(hr) &  $\sigma$(m\,s$^{-1}$)  &  Date  &  Nbr  & T(hr) & $\sigma$(m\,s$^{-1}$) \\
            \noalign{\smallskip}
            \hline
01/27 & 232 & 5.56 & 5.64 &       &     &     &       \\
01/28 & 214 & 5.51 & 2.76 & 01/29 &  92 & 3.84 & 5.21 \\
01/29 & 233 & 5.57 & 2.70 & 01/30 & 155 & 6.15 & 4.00 \\
01/30 & 282 & 8.30 & 3.18 & 01/31 & 171 & 6.79 & 4.21 \\
01/31 & 287 & 8.44 & 2.71 & 02/01 & 166 & 6.59 & 3.90 \\
02/01 & 354 & 8.45 & 2.70 & 02/02 & 194 & 7.78 & 3.84 \\
02/02 & 389 & 9.28 & 2.40 & 02/03 & 196 & 7.79 & 4.49 \\
02/03 & 266 & 7.82 & 2.75 & 02/04 & 153 & 6.08 & 4.02 \\
02/05 & 336 & 7.93 & 3.43 &       &     &      &      \\
02/06 & 280 & 8.08 & 3.00 &       &     &      &      \\
02/07 & 269 & 8.01 & 2.87 &       &     &      &      \\
02/09 & 101 & 3.00 & 3.08 &       &     &      &      \\
            \noalign{\smallskip}
            \hline
         \end{tabular}
\caption {\label{tab_1} Observation log (UT)}
\end{table}

   \begin{figure}
\vspace{0cm}
\resizebox{\hsize}{!}{\includegraphics{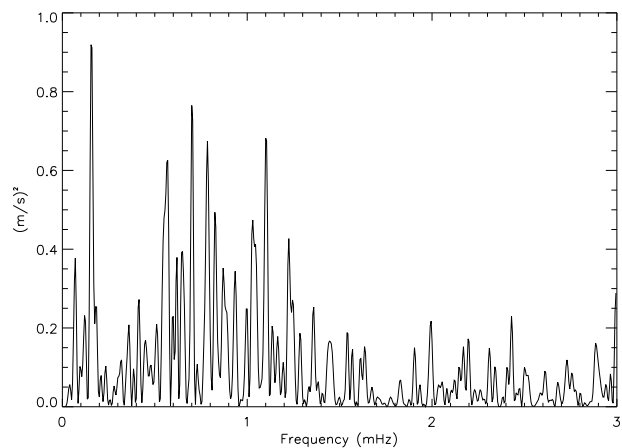}}
\vspace{0cm}
      \caption[]{Power spectrum of Procyon A from
Doppler shift measurements over 15 hours, with ELODIE and AFOE.}
         \label{Fig_proc1}
   \end{figure}
 
\begin{figure*}
  \centering
\begin{minipage}[b]{0.5\textwidth}
\centerline{\footnotesize ELODIE 1999}
\scalebox{0.5}{\includegraphics{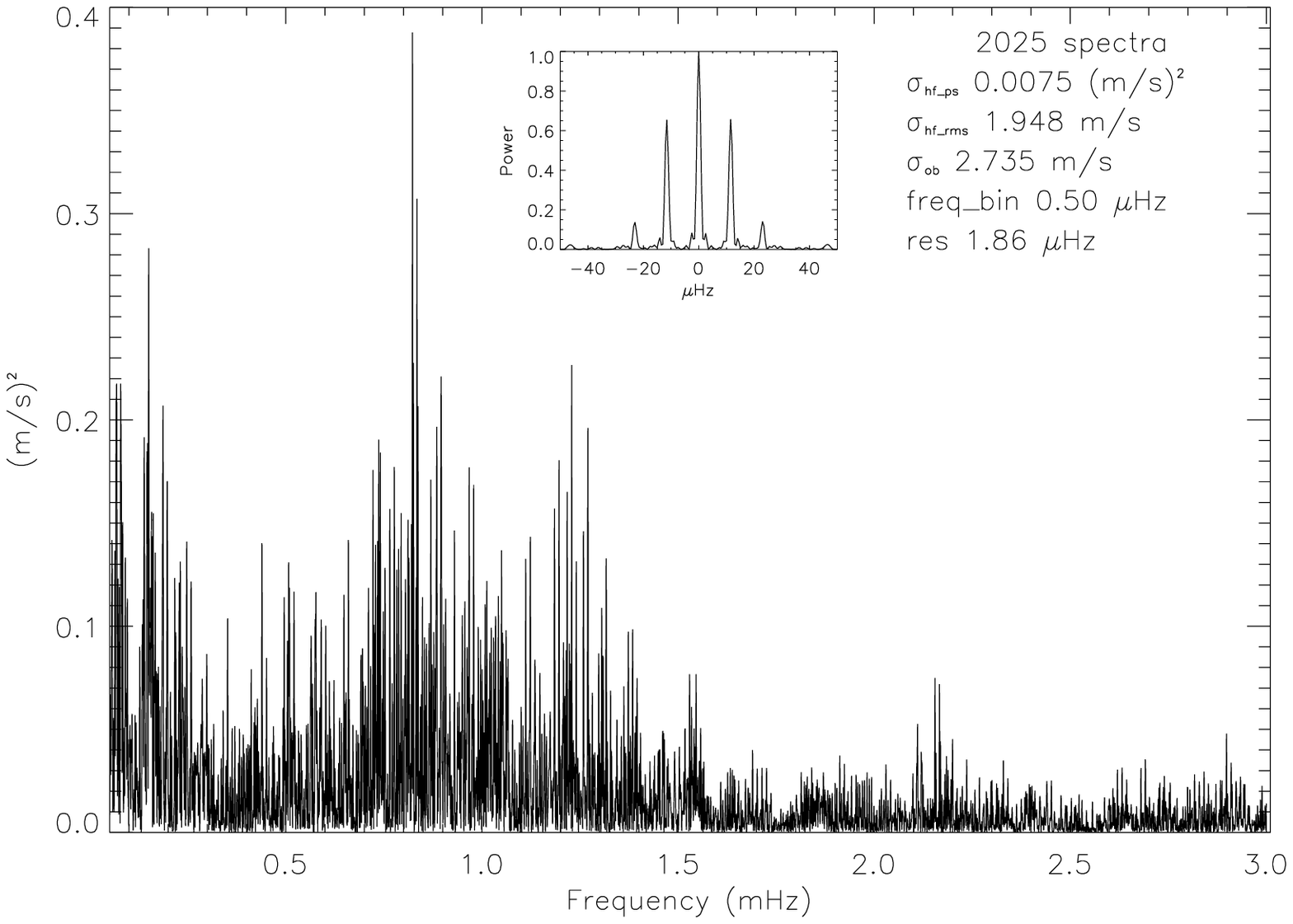}}
\end{minipage}%
\begin{minipage}[b]{0.5\textwidth}
\centerline{\footnotesize AFOE 1999}
\scalebox{0.5}{\includegraphics{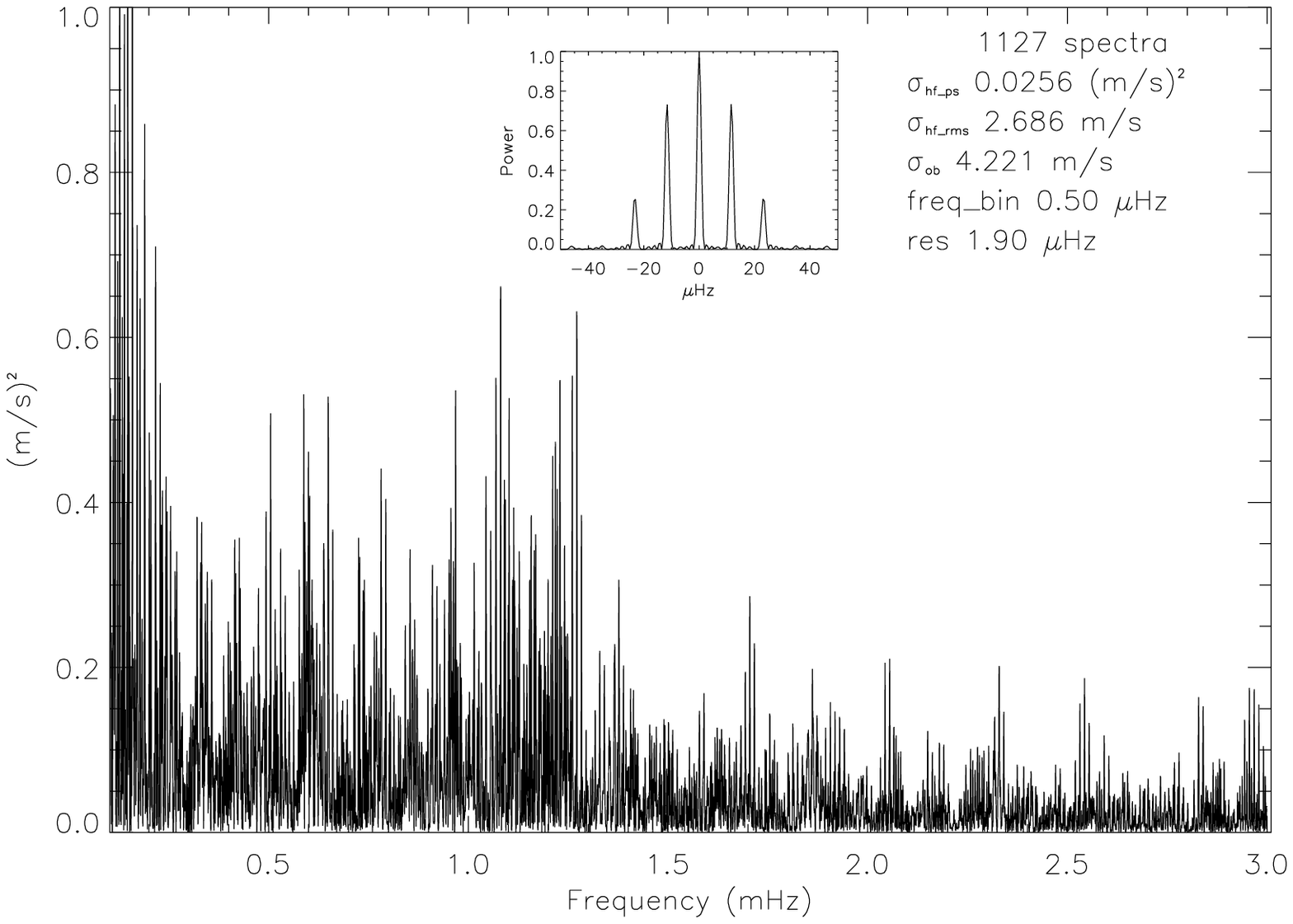}}
\end{minipage}

\vspace{0.5cm}

\begin{minipage}[b]{0.5\textwidth}
\centerline{\footnotesize ELODIE/AFOE 1999}
\scalebox{0.5}{\includegraphics{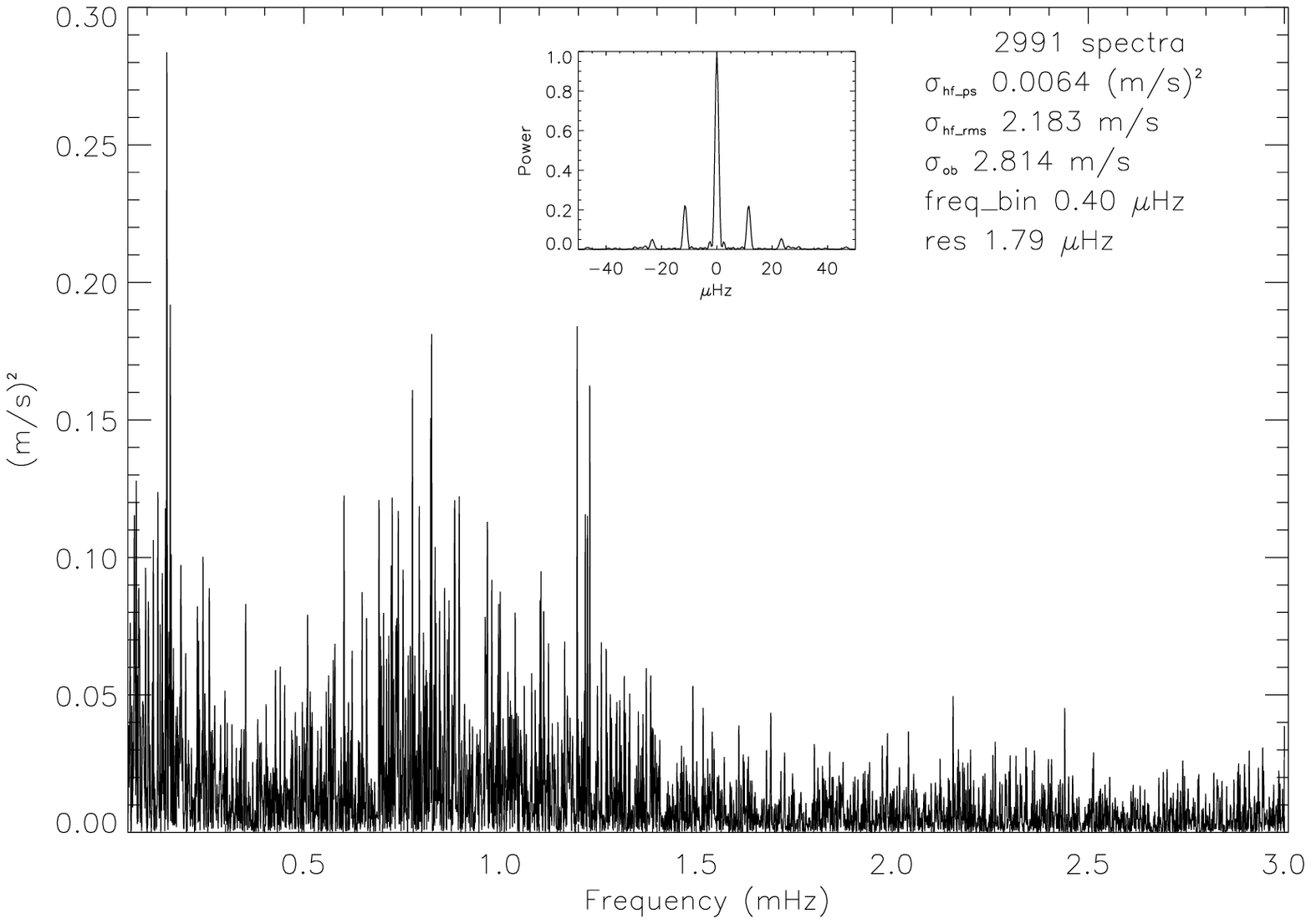}}
\end{minipage}%
\begin{minipage}[b]{0.5\textwidth}
\centerline{\footnotesize Simulations (ELODIE/AFOE) }
\scalebox{0.50}{\includegraphics{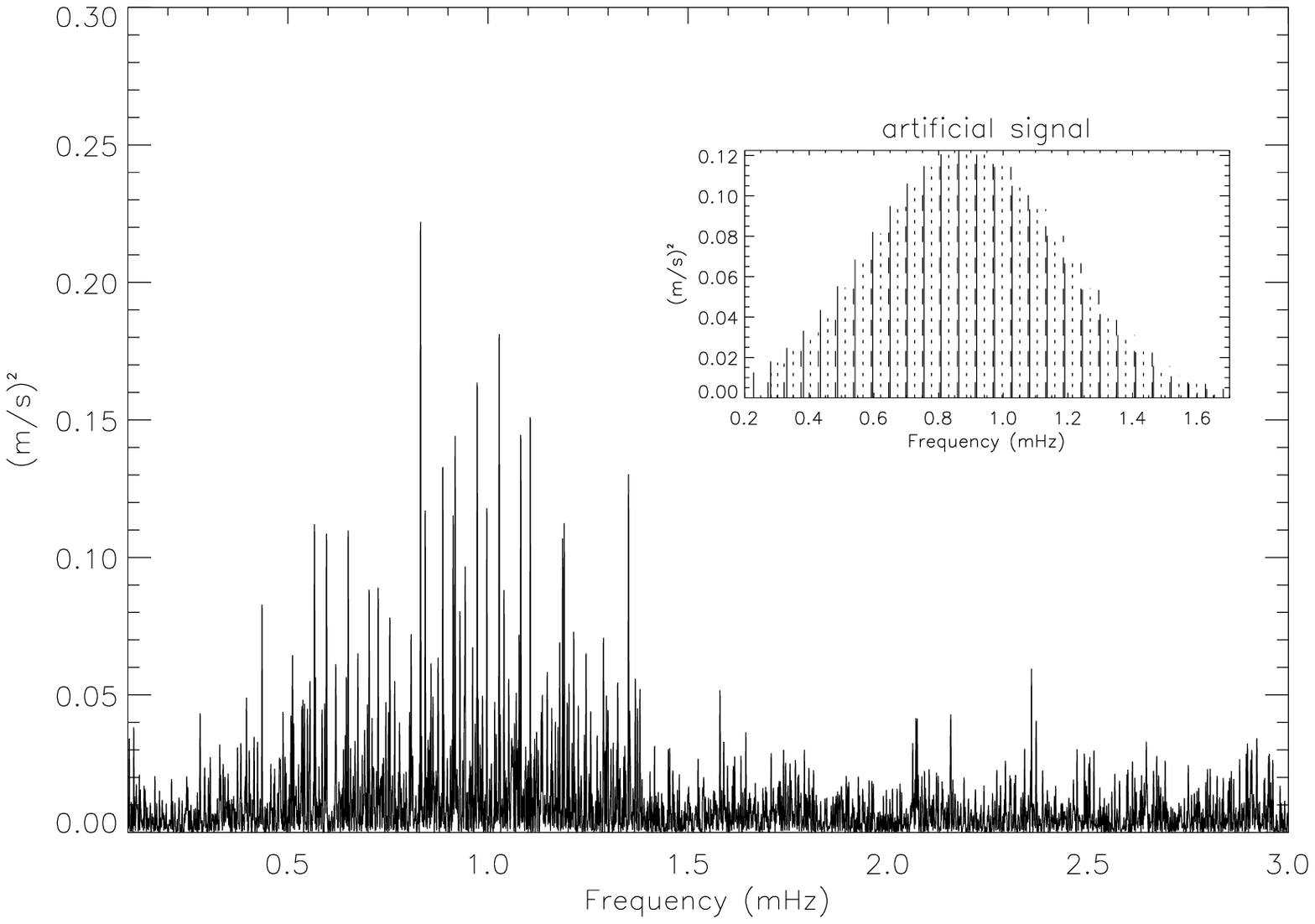}}
\end{minipage}
      \caption[]{
Top panels: Power spectra of the ELODIE and AFOE Doppler shift measurements (7 nights) of Procyon. Bottom left panel: Power spectrum of the combined data sets. Bottom right panel: Simulated spectrum computed with  noise and with the same observation times as the combined sequence ELODIE/AFOE.}
         \label{Figzetaher4fig}
\label{4Fig_proc}
   \end{figure*}

Due to variable weather conditions and lower mean S/N than in our 1998 observations, we de-correlate the Doppler
 velocity from residual flux variations
  in and between the orders, used for the calculation of the final weighted mean Doppler signal.

A journal of Procyon A observations in 1999 is given in 
Table~\ref{tab_1}. 
The table lists the length of the individual runs in hours, the number of selected exposures and the rms scatter of the time series.  
To illustrate the precision of each data-set, we show in Fig.~\ref{Fig_obs}
the resulting Doppler signal from both sites. 

\section{Power spectra analysis}
\label{section_period}

The power spectrum of the weighted mean Doppler shift measurements for one of the best nights (high S/N, low rms residual)
 of Procyon observations is presented in Fig.~\ref{Fig_proc1}. 

 Inspection of the other single-night spectra confirms the presence of an excess power
 and several common peaks that modulate the envelope between 0.3\,mHz and 1.7\,mHz. Comparison
 of individual power spectra allows us to search for the repetitive occurrence
of peaks, free from aliasing, from one night to an another, but the frequency resolution is too low to distinguish the mode fine structure.
We find in a few cases large peaks which highly exceed the S/N in the region
 of the hump of power. This characteristic could be explained as 'large excitations' from the studies of the solar \textit{p}-mode spectrum \citep{Elsworth95}.
 In the solar case, the lifetime of the strongest modes are only of the order
    of a few days and their evolution can be followed on a short time scale when their frequencies are visibly
separated and free from mutual interferences. In our observations the occasionally large excitations stay visible in the power spectrum from three or four successive nights but more or less important interferences with noise have also to be
considered. In general, the relative amplitudes of the frequencies vary from night to night, with the result that a given frequency
 may dominate the amplitude spectrum in some data segments and be entirely undetectable in the others.
To address these issues, we performed many frequency analyses of the entire data set, numerous subsets of 
the data and simulated time series having the same sampling function as the observations.

 In Fig.~\ref{4Fig_proc} (upper panels), we show the power spectra from ELODIE and AFOE one-site measurements  of Procyon A over seven consecutive nights. Note that the observing run with the ELODIE/OHP spectrograph was longer, twelve nights \citep[see periodogram in][]{Martic01b}. 
The power spectra were calculated using the Lomb-Scargle (LS) modified algorithm \citep{Lomb76,Scargle82}
with statistical weights. The mean white noise level ($\sigma_{\rm hf\_ps}$), the rms scatter ($\sigma_{\rm ob}$),
 the frequency resolution and computation bin are indicated for each data set.  The noise level is higher in the AFOE data, which makes less evident the excess of power in the low-frequency part of the spectra. 

The time series from the two sites were merged by first subtracting a second order polynomial fit
 and the clipping data over $2\,\sigma$ for each night. This process yields similar rms scatter of data from one night to another and allows us to compute the power spectrum of combined time series without using weights (see bottom left panel of Fig.~\ref{4Fig_proc}). To verify that the amplitude of the excess signal was not severely reduced by removing outlying points over $2\,\sigma$, we calculated also the power spectrum using statistical weights (1/$\sigma^2$) of the time sequence after $4\,\sigma$ clipping. By comparing these two power spectra, we note that the frequencies of the major peaks remain unchanged but with more power in the sidelobes of the weighted spectrum. For these data the application of the weights has not produced an important reduction in noise but rather degraded the window function.

 One can see in Fig.~\ref{4Fig_proc} that the overall shape of the excess power around 1\,mHz is not symmetric and does not have the typical bell-like envelope of the solar oscillations.
 This can be explained by the effects of gaps/noise and amplitude modulation that are intrinsic
  to the star (beating between pairs of modes that are closely spaced in frequency and finite
   lifetime of each oscillating mode). The effects of noise interference and the window function
    are visible in the power spectrum (see right lower panel of Fig.~\ref{4Fig_proc})
     of the simulated time series with an artificial oscillating signal
 and the rms almost
   equal to that of the observed data sets. 

We performed several simulations
in the presence of noise with the oscillation signal based on frequencies identified in section~\ref{section_modes}. The amplitude distribution of the 
generated signals was produced as in \citet{Barban99}
but with mode ($l,m$) ratios computed for updated Procyon parameters
 and with the maximum mode amplitude of about 35\,cm/s near the peak of the excess power.
 
 In the periodogram of the joint ELODIE and AFOE observations
 several isolated peaks are present between 0.3\,mHz and 1.4\,mHz, due to a much better window function
 (see the amplitude of the sidelobes in the inset of Fig.~\ref{4Fig_proc}). These frequencies were used as starting ones for the comb power
 analysis and asymptotic fit calculations (cf. next section).

\section{Echelle diagrams}
\label{section_echdiag}

In the solar case, the asymptotic theory \citep[see e.g.,][]{Tassoul80} provides quite good results
 for the acoustic modes with high radial order $n$ and low degree $l$. The asymptotic relation is expected to hold for the solar-like $p$-mode oscillations in other stars and gives mode frequencies as
 \begin{eqnarray} \label{eq_1}
\nu({nl})=\Delta\nu(n+\frac{1}{2}l+\frac{1}{4}+\alpha)+\epsilon_{nl}
\end{eqnarray}
where $\Delta\nu$ is the inverse of twice the sound travel time between the centre and the surface, $\alpha$ is dependent on the reflective properties of the surface and $\epsilon_{nl}$ is a small correction term sensitive to the central structure of the star.

In the vicinity of some radial order $n_0$ one can parametrize the expected mode frequencies as
\begin{eqnarray} \label{eq_2}
\nu(n_\mathrm{p},l)\simeq\nu_{0}+\Delta\nu_{0}(n_\mathrm{p}+\frac{1}{2}l)-l(l+1)D_{\rm 0}
\end{eqnarray}
where $n_\mathrm{p}=n-n_0$, $\nu_{0}$ is the frequency with $n=n_0$ and $l=0$, $\Delta\nu_{0}$ represents the so-called mean large frequency separation
between $p$-modes of same degree and adjacent $n$. $D_{\rm 0}$ is related to the so-called small frequency separation $\delta\nu_{l,l+2}$ where 
$\delta\nu_{l,l+2}=\nu(n,l)-\nu(n-1,l+2)\simeq(4l+6)D_{\rm 0}$.

The third separation $\delta\nu_{0,1}$, which represents the amount by which $l$=1 modes are offset from the midpoint between the $l$=0 modes on either side \citep[see e.g.,][]{BeddingKjeldsen03} can be defined as $\delta\nu_{0,1}=\nu(n,0)-\frac{1}{2}(\nu(n,1)+\nu(n-1,1))$. If the asymptotic relation holds exactly,
  $D_{\rm 0}=\frac{1}{6}\delta\nu_{02}=\frac{1}{10}\delta\nu_{13}=\frac{1}{2}\delta\nu_{01}$. 

To the first order, a star pulsating with a number of consecutive radial modes of a given $l$ has a comb-like eigenspectrum with a spacing $\Delta\nu_{0}$ and if both even and odd $l$ modes are present, with spacing $\Delta\nu_{0}/2$. To estimate $\Delta\nu_{0}$ in the region of excess power of
 the Procyon frequency spectrum we used the comb response method.

 In previous analysis of the single-site data \citep{Martic01a} we applied the comb analysis 
to the CLEAN-ed power spectra because of important  aliases
 at spacings $\pm\,11.57\,\mu$Hz from the largest peaks.
 To improve the accuracy of these results and to see
 whether we can use the comb response (CR) to distinguish the even from odd $l$ mode frequencies, we rewrite the function \citep[cf. Eq.~2 in][]{Kjeldsen95} as 

   \begin{figure}
\vspace{0cm}
\resizebox{\hsize}{!}{\includegraphics{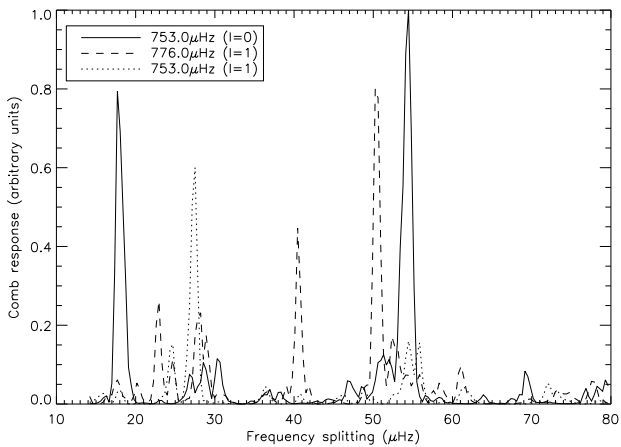}}
\vspace{0cm}
      \caption[]{Comparison between modified comb responses
computed for two central frequencies of the power spectrum of Procyon data
 (see description in text).}
         \label{Fig_comb}
   \end{figure}

\begin{eqnarray} \label{eq_3}
C(\nu_\mathrm{0},\Delta\nu_\mathrm{0})=\prod_{i=0}^{4}\left[\,S(\nu_\mathrm{0}+i\frac{\Delta\nu_\mathrm{0}}{2}+\delta)\,S(\nu_\mathrm{0}-i\frac{\Delta\nu_\mathrm{0}}{2}+\delta)\right]^{\alpha_i}
\end{eqnarray}
 \\
where
 $\delta=\pm(i\bmod 2)D_0$ according to mode degree $l=0$ or $l=1$ at the
 central frequency $\nu_{0}$
and $\alpha_i=1$ for $i=1$ or 2,
 $\alpha_i=0.5$ for $i=3$ or 4. 

\begin{figure}
  \centering
\begin{minipage}[b]{0.5\textwidth}
\scalebox{0.5}{\includegraphics{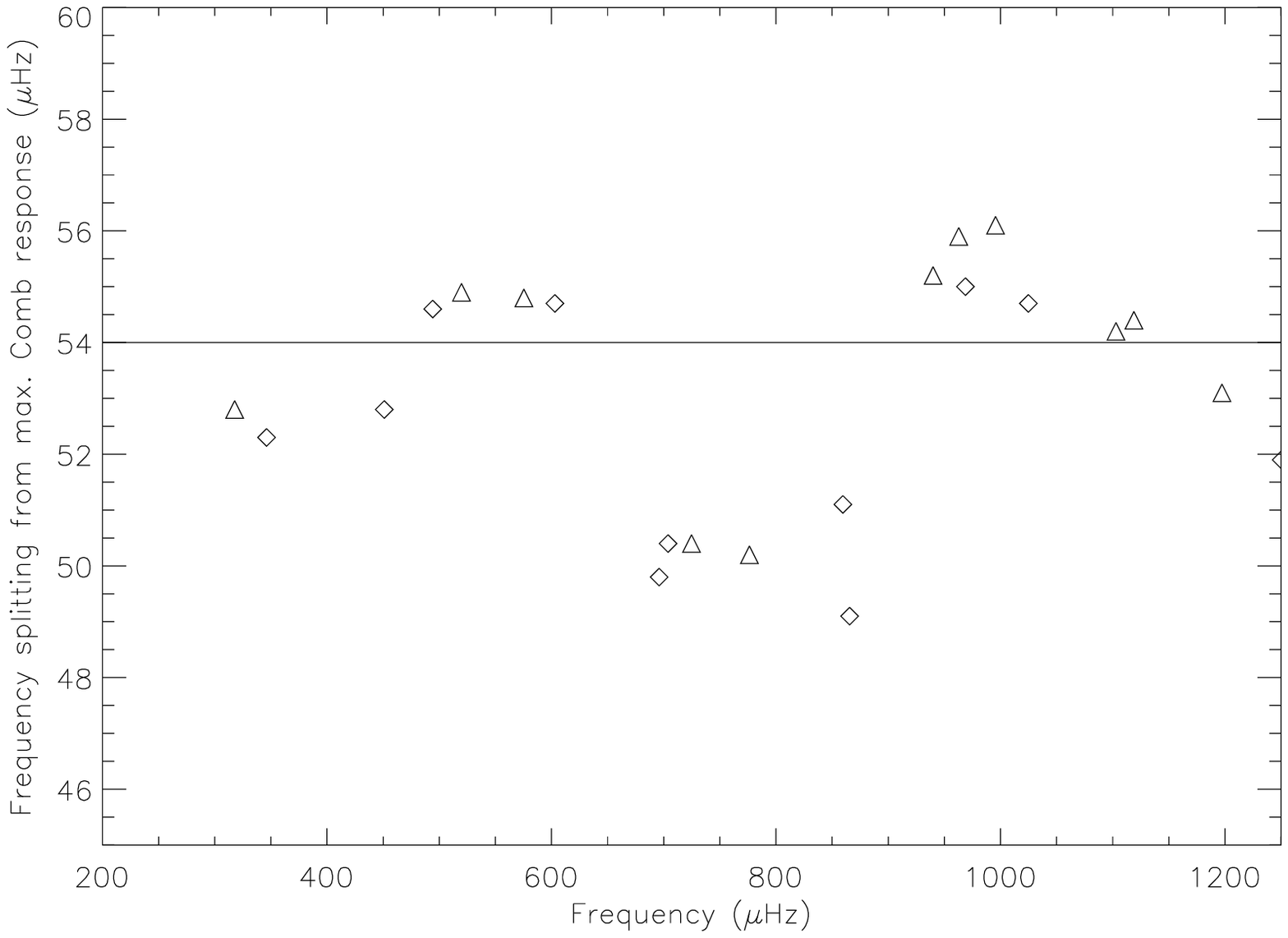}}
\end{minipage}
\vspace{0.5cm}
\begin{minipage}[b]{0.5\textwidth}
\scalebox{0.5}{\includegraphics{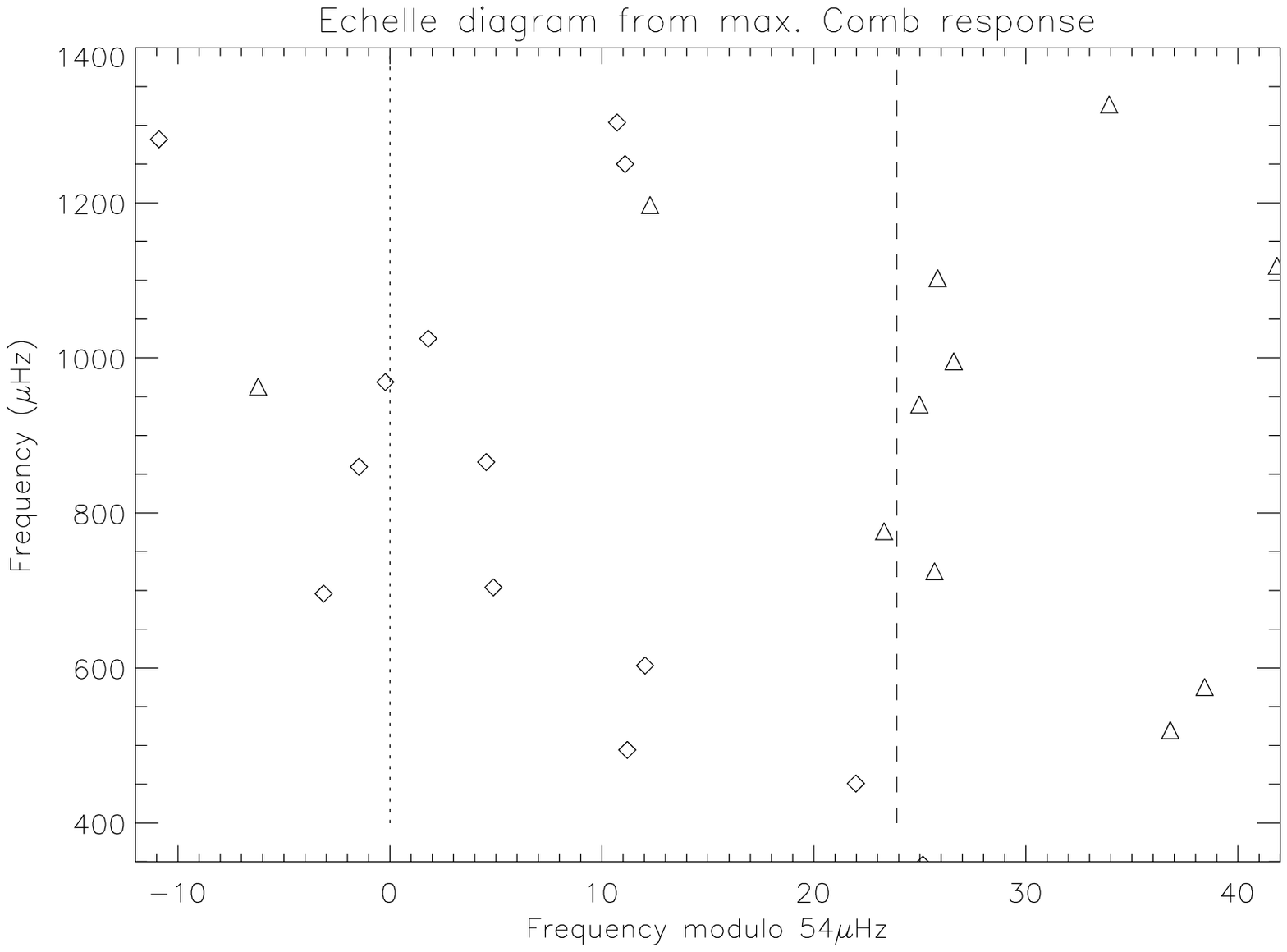}}
\end{minipage}
      \caption[]{
Top panel: The first-order spacings over several frequency ranges, as determined from comb analysis. Bottom panel: Echelle diagram of the selected frequencies with the best comb responses ($l=0, 1$ indicated respectively by diamonds and triangles).}
\label{Fig_multicombech}
   \end{figure}
%

   \begin{figure*}
\vspace{0cm}
\resizebox{\hsize}{!}{\includegraphics{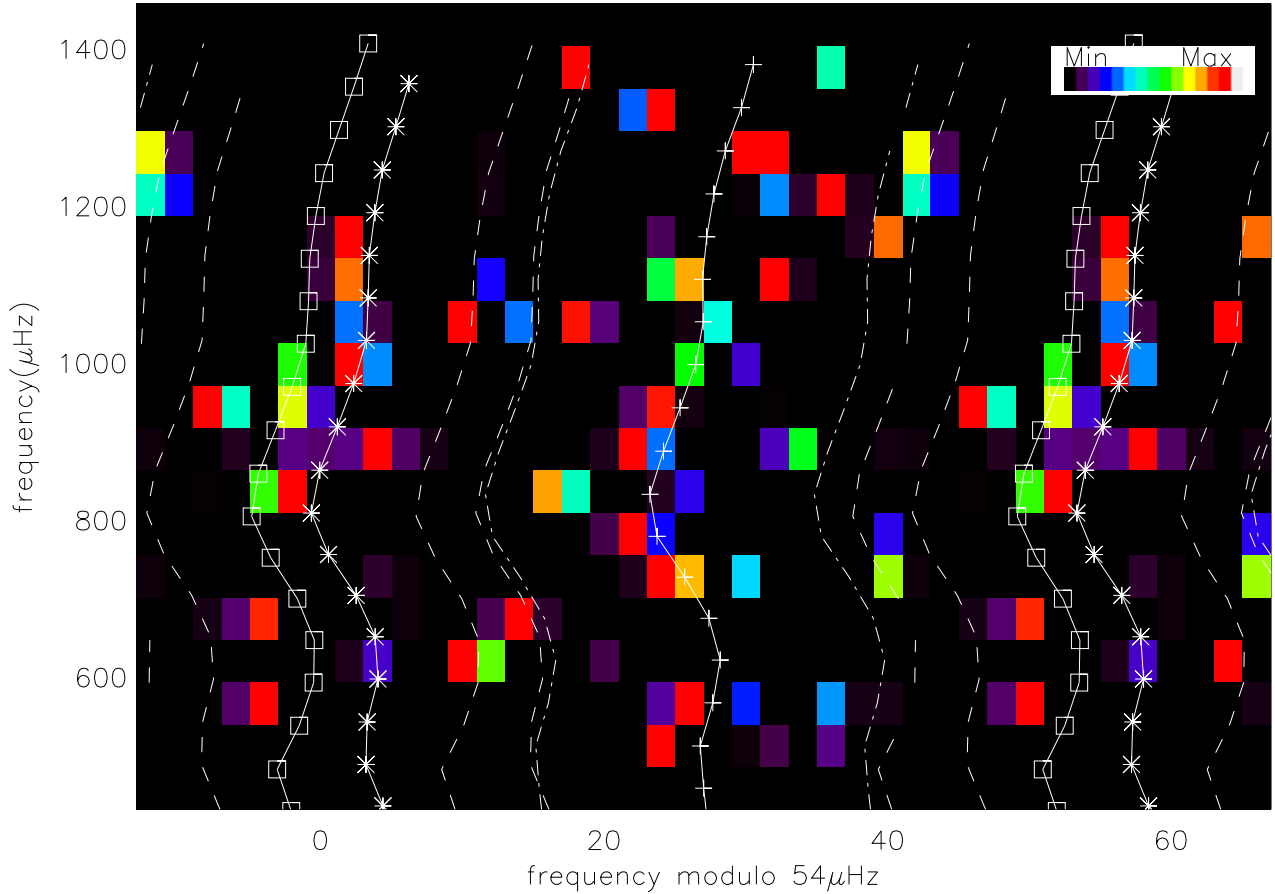}}
\vspace{0cm}
      \caption[]{Example of an echelle diagram from Procyon amplitude spectrum of the joint
ELODIE and AFOE radial velocity measurements, computed for the modulo mean $\Delta\nu_{\rm 0}$= 54\,$\mu$Hz. The frequencies from the standard model \citep[see][]{Chaboyer99} are indicated by asterisks, plus signs and squares respectively for $l=0,1,2$, dashed lines are at $l=0,1,2 \pm\,11.6\mu$Hz).
The inset shows the colour coding applied to the values in each succesive folding frequency range.}
         \label{Fig_imagech}
   \end{figure*}

 The modified comb responses (Eq.~\ref{eq_3}) were then calculated
 at the frequencies of the highest peaks in the power spectrum. 
 For each central frequency $\nu_{0c}$, alternatively considered
 as $l=0$ and $l=1$,
 we have searched for the maximum CR
 for \mbox{$10\,\leq \Delta\nu_{0}\leq 80 \,\mu$Hz} and 
$0.3\,\leq D_{\rm 0}\leq 2$.  

In Fig.~\ref{Fig_comb} we show examples of the modified comb response computed for two $\nu_{0c}$ frequencies at 753\,$\mu$Hz and 776\,$\mu$Hz. 
The responses are scaled to the maximum peak obtained at
$53.5\,\mu$Hz when $\nu_{0c}$= 753\,$\mu$Hz was assumed to be $l=0$.

From the initial set of the highest peaks corresponding to the modes, but also to the aliases, the 
comb analysis gives the reduced set of $\nu_{0c}$ frequencies and corresponding $\Delta\nu_{0c}$ for which the maximum CR is obtained. The upper panel of 
Fig.~\ref{Fig_multicombech} shows  the variation of the first-order spacing over several different frequency ranges, as determined from comb analysis. The arithmetic mean of these spacings is $54\,\mu$Hz, a value for a large separation in good agreement with the theoretical predictions from models \citep{Chaboyer99,dimauro01,Provost02}. In addition, this exploration allowed us, from the comparison of the relative power of the best responses, to separate temporarily
 the modes by degree. In Fig.~\ref{Fig_multicombech} (lower panel), we show these frequencies in the \textit{echelle} diagram \citep[cf.][]{Grec83} for the modulo mean $\Delta\nu_{\rm 0}$= 54\,$\mu$Hz. 

In a second step we calculate the echelle diagram so as to preserve the amplitude information.
It can be constructed either from the power or CLEAN-ed spectra by adding (within cells of the
order of the frequency bin) the values modulo large splitting, over threshold determined by the mean noise level of a time sequence. This means that any points lower than the noise level are set at this level. 
The echelle diagrams constructed in this way are used to compare the overall distribution of the power for different possible average splittings. In general, the ridge like structure appears for the mean $\Delta\nu_{\rm 0}$ determined from previous comb-like analyses. This method was first tested on solar data obtained with the same instrument during a daytime run. The echelle diagram of the single-site ELODIE observations were shown in \citet{Martic01b} and discussed in \citet{Provost02}.

We should point out that the distribution of
the power in these kinds of echelle representations depends greatly on the 
starting folding frequency $\nu_0$ and the threshold level, especially when mode amplitudes are decreased by destructive
 interference in some frequency ranges. In Fig.~\ref{Fig_imagech} we show an example of such an echelle diagram based on the ELODIE and AFOE joint observations. To increase the visibility of the modes
  in this image we modified the power spectrum by a simple normalization to the highest peak in each successive
   folding frequency range. The maximum power
is concentrated in the ridge that is identified as $l=1$ and which shows the offset from $0.5\Delta\nu_{0}$ between the $l=0$ modes. The departure from a linear asymptotic relation is evident.
  It explains why, for the stars like Procyon, classical methods used for mode diagnostics
   are not easily applicable \citep[cf.][]{Bedding01}.

\section{Mode identification}

\label{section_modes}
   \begin{figure}
\vspace{0cm}
\resizebox{\hsize}{!}{\includegraphics{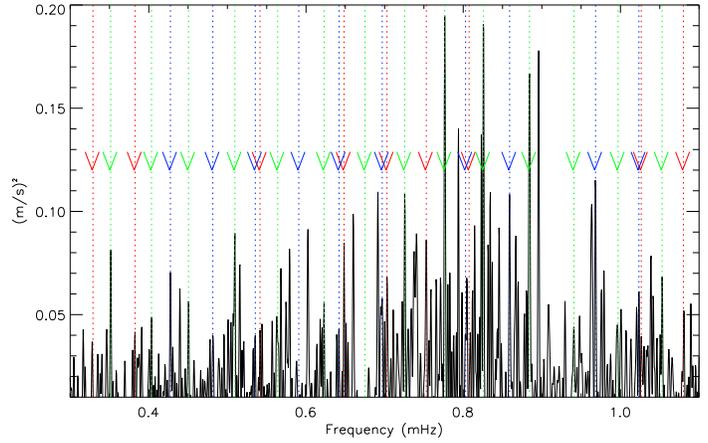}}
\vspace{0cm}
      \caption[]{Identification of the $p$-modes in the region
 of the excess power of Procyon A. Dashed lines in blue, red and green correspond
 respectively to modes $l=2,0,1$.}
         \label{spec_freq}
   \end{figure}

The best values, based on our modified comb analysis, for $\nu_{0c}$,
 $\Delta\nu_{0c}$ and $D_{0c}$,
 were then used to reconstruct the parameterized frequencies in separate frequency regions ($\nu_{0c}\pm\,2\Delta\nu_{0c}$). These 'model' frequencies have a significant
departure from the asymptotic relation for one set of mean fit parameters. From this step, we have proceeded in an iterative manner to identify the modes.

The observed high amplitude frequencies were classified by degree comparing to the model ones and used to obtain new fit parameters by performing multidimensional minimization of Eq. \ref{eq_2}. This method was used because we first selected the peaks with the strongest
amplitudes, which means that successive $n_{\rm p}$, or \textit{l} pairs of modes are not always present. We calculated the fit for various combinations of modes with different degree, using the downhill simplex method \citep{NelderMead65}. A certain number of high amplitude frequencies were then shifted by $\pm\,11.57\,\mu$Hz and included in the selection of the modes, which improved the convergence (small max error). In Fig.~\ref{spec_freq}, we show the selection of frequencies in the region of excess power at $4\,\sigma$ above the mean white noise, for which we obtained the best fit. The values for $\Delta\nu_0$ and $D_0$ from the fit are~:

$\Delta\nu_0=53.6\,\mu$Hz  

$D_0=0.85\,\mu$Hz, $\delta\nu_{02}\approx6D_0=5.1\,\mu$Hz

Given the complexity of this spectrum, the noise background, the short mode lifetimes \citep[cf.][]{Brown91} and the possibility of  cross-talk between a real frequency and an alias etc., we checked the presence of this set of frequencies in other independent subsets of the data. We calculated periodograms from the sequences of three and four consecutive nights which have
a sufficient resolution ($4\,\mu$Hz) to resolve $l=0, 2$ from $l=1$ modes. The independent short time series, chosen for their low rms, were processed using 
CLEAN \citep{roberts87} and CLEAN-est \citep{Foster95} algorithms to minimize the influence of sidelobes in the power spectra. The advantage of the second one is that it gives directly an initial set of frequencies that can be used for the mode identification. Deconvolution methods allow for the identification of modes with smaller amplitudes in the spectrum but isolate also the aliases instead of the modes. Unfortunately, in the case of
 Procyon the frequency aliases between $l=0$ and $l=1$ are very close to each other, i.e. $\nu_{l=0}+11.57 \approx \nu_{l=1}-11.57$. 
We should also note that, whichever subset is used, there are gaps in the amplitude spectrum where frequencies in the comb pattern are not excited to observable amplitudes. 

In \citet{Martic01a} we showed the histogram of the major peaks that were considered as recurrent when present within $\sim\,5\,\mu$Hz from one sequence to an another using single-site data. From a complementary search of the modes in short time sequences, we obtained a set of recurrent frequencies for further mode analysis of longer sequences where finite lifetime and random phases from the stochastic nature of mode excitation interfere and may reduce the observed amplitudes of modes.

\begin{figure}
\vspace{0cm}
\resizebox{\hsize}{!}{\includegraphics{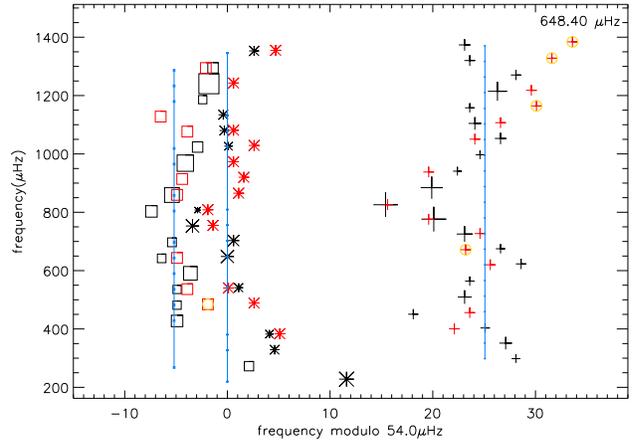}}
\vspace{0cm}
      \caption[]{Echelle diagram of the identified modes of
degrees $l=0,1,2$ from two-site joint Procyon observations (cf. Table~\ref{tab_freq}). For comparison, we indicate with the symbols in red the frequencies detected from previous OHP/ELODIE 1998 run. 
Vertical lines correspond to the best fit of the observed frequencies to the 
asymptotic relation, with $\nu_0$=648.4\,$\mu$Hz, $\Delta\nu_0=53.6\,\mu$Hz and $D_0=0.85\,\mu$Hz}
         \label{fig_freq1}
   \end{figure}

Finally, we explored all the peaks at $2.5\,\sigma$ above the mean white noise level (7\,cm\,s$^{-1}$) in the amplitude spectrum of
the time sequence from coordinated two-site measurements. The identified modes, with the assigned $l$-values ($l=2, 0, 1 $ marked respectively with squares, asterisks, and plus signs) are presented in Fig.~\ref{fig_freq1}, with 
the size of the symbols scaled to the amplitude of the maximum peak in the spectrum.  

For comparison we also show modes (with red symbols) from the 1998 observing run (cf. Paper I). Since, the aliases are more important in these single site measurements, we used the CLEAN  algorithm in the data reduction. In some cases, the amplitude of the peak is substituted by the higher one of the corresponding alias (symbols in circles). The overall distribution of the modes confirms and completes the results from two-site observations, which increases the confidence in these mode frequencies. One can see in Fig.~\ref{fig_freq1} that the $l=1$ mode distributions are consistent between two sequences while the $l=0, 2$ modes are scattered in the range of the $2\,\mu$Hz error bar. For a given frequency resolution, the uncertainty of the $l=0, 2$ mode identification is due to the decrease of the small separation $\delta\nu_{02}$ with the frequency.

\begin{table}
\caption {Mode frequencies of Procyon A (in $\mu$Hz)}
\label{tab_freq}
   \begin{center}
\begin{tabular}{|l|lll|}
            \hline
            \noalign{\smallskip}
$n_{\rm p}$   &  $l=2\,(n_{\rm p}\,-\,1)$   &   $l=0$       &  $l=1$         \\
            \noalign{\smallskip}
            \hline
-10 & 272.5 (273.9)  &  (278.)    &  299.    \\
-9  &                &  329.    &  352.    \\
-8  &                &  382.    &  (400.8) 403.5    \\
-7  & 427.5          &          &   450.5 (456.6)          \\
-6  & 481.5          &  (484.)  &  (507.6) 509.7      \\
-5  & 535.5          &  541.3   &  564.6  (568.)   \\
-4  & 590.8          &  (596.)  &  (619.)  623.     \\
-3  & 642.           &  648.4   &  (671.6) 675.      \\
-2  & 697.           &  702.6   &  725.5   (726.3)          \\
-1  & (749.6)          &  753.    &  776.7        \\
n0  & 803.           &  807.3   &  825.5 or 834.5       \\
+1  & 859.1          &         &   884.5          \\
+2  & (908. 910.)    &         &   941.            \\
+3  & (964.) 968.    &         &   997.          \\
+4  & (1021.) 1023.5 & 1026.5  &  1050.5 (1052.3)           \\
+5  & (1076.5)       & 1080.  &  1104.5        \\
+6  & (1130.)        & 1134.   &  1158.       \\
+7  & 1186.          & (1190.) &  (1214.5) \\
+8  & 1241.          & (1244.6)&  1270.    \\
+9  & 1295.          &         &  1320.8 (1328.)   \\
+10 &                & 1353.   &  1373.  (1384.)       \\
            \noalign{\smallskip} 
            \hline
         \end{tabular}
   \end{center}
\end{table}

We list in Table~\ref{tab_freq} the frequencies of identified modes, detected more than once in different data sets. Some of the frequencies,
 marked in parenthesis, are detected 
in longer sequences with better resolution but with a more complicated window function (see Table~\ref{tab_1}). In higher frequency ranges, two pairs of peaks 
    (1320.8, 1373 and 1328, 1384 $\mu$Hz)
     are recurrent in different subsets of the data at the $3\,\sigma$ level above the noise and can be equally identified as $l=1$ modes or aliases. Note that at the frequencies of the second pair, the modes would follow an important curvature in the echelle diagram.

\mathversion{bold}
\section{$p$-mode analysis}
\mathversion{normal}
\label{p-mode analysis}
 
Different combinations of frequencies of low degree $p$-modes carry information on different parts the of the stellar structure. The large separation between frequencies of modes of given degree $l$ and consecutive radial orders is sensitive to the global structure and depends on the outer layers of the star. It is mainly related to the stellar radius
or the dynamical frequency $\propto(GM/R^3)^{0.5}$. Mean separations between frequencies of modes with degree, $l, l +2$ or $l, l +1$, which penetrate differently in the central layers, are mainly sensitive to the structure of the deep interior and hence to the evolutionary state. From different models of Procyon by \citet{Provost02}, in the range of considered stellar parameters and overshoot parameters, the mean small spacings $\delta\nu_{02}$, $\delta\nu_{01}$, $\delta\nu_{13}$ vary respectively within $1\,\mu$Hz, $1.5\,\mu$Hz and $2\,\mu$Hz. Such small differences are difficult to measure with the precision of our data.

   \begin{figure}
\vspace{0cm}
\resizebox{\hsize}{!}{\includegraphics{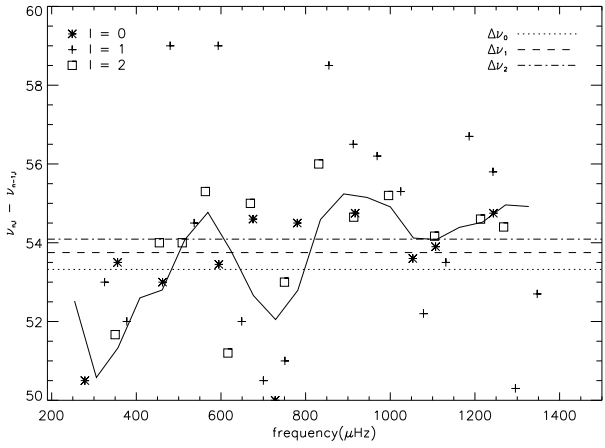}}
\vspace{0cm}
      \caption[]{Variation of the large spacing between the modes of same degree,
compared to the predicted curve from the standard model for $l=0$
 \citep{Chaboyer99}. The observed average large separations for $l=0,1,2$
 degree are indicated by the horizontal lines.
}
         \label{fig_delt_nu}
   \end{figure}

In Fig.~\ref{fig_delt_nu}, we show the large separation as a function of frequency,
 for $l=0, 1, 2$. Note that the mean large spacing does not depend much on the degree. When the successive orders are not detected,
  the separation was calculated by linear interpolation. Few mode frequencies are dispersed more than $\pm\,2\,\mu$Hz around the mean separations. 
  One of the strongest peaks at $825\pm\,2\,\mu$Hz, identified as the $l=1$ mode is eventually displaced by an avoided crossing
   \citep[see e.g.,][]{Audard95,dimauro03}, but additional data are needed to confirm this possibility.

Considering the small spacing of about only $4\,\mu$Hz above 0.7\,mHz, it is rather difficult with the present data to distinguish without ambiguity $l=0$ from $l=2$ modes of successive radial order. We should however note that the number of identified $l=2$ modes is higher than for $l=0$. The observed larger amplitude of these modes is probably due to interferences between modes of different $m$. To estimate the rotational splitting of the modes we assumed uniform rotation and the inclination of the rotational axis equal to the
inclination of the orbital axis of the binary system ($31.1^{\circ}$) determined by \citet{Girard00}. 
With an estimated radius of $2.071\,R_{\odot}$ and $v$sin$i$ between 
2.7\,km/s \citep{Prieto02} and 6.1\,km/s \citep{Pijpers03} the corresponding rotational splitting is respectively 0.58 and $1.3\,\mu$Hz. We performed several simulations (cf. section~\ref{section_period}) in which synthetic oscillation spectra are computed with the rotational frequency between these two values. As verified with simulations for a given rotational splitting a pair of $l=0, 2$ modes may overlap and produce one single large peak at shifted frequency in the power spectrum. The frequency resolution
of joint OHP and AFOE data is not sufficient to attempt a determination of rotational splitting.

\section{Oscillation amplitudes}

Based on Christensen-Dalsgaard \& Frandsen (1983) computations of expected amplitudes from stochastic excitation in stars on or near the main sequence, \citet{KjeldsenBedding95} proposed a scaling for the oscillation velocity amplitude  
\mbox{$V_{\rm osc}\propto\,L/M$} , where \textit{L} is the luminosity and \textit{M} the mass of the star. The calculations by \citet{Houdek99} confirmed this result, although they fitted better the velocity with $V_{osc}\propto\,(L/M)^s$, and $s\sim1.5$. However, both the simple scaling law and the computation of Houdek et al. grossly overestimate the observed amplitudes of Procyon ($\lesssim\,50$\,cm/s).
 
To reduce the "expected" amplitude of Procyon, \citet{KjeldsenBedding01}
suggested that $V_{\rm osc}$ is independent of the effective temperature amongst stars of a given $M$ and radius $R$. They proposed 
$V/V_{\sun}\propto\,g_{\sun}/g$  (\textit{g} surface gravity) 
which, for Procyon, yields lower velocity ($70$\,cm\,s$^{-1}$) than the scaling $V/V_{\sun}=2[(g_{\sun}/g)^{0.6}+(g_{\sun}/g)^{4.5}]^{-1}$  \citep{Gilliland93}, but still higher than the observations. 

Theoretical amplitudes from models are also overestimated, almost three times higher than the
observed one. It appears \citep{HoudekGough02}, therefore, that there is something wrong with the
theory of the pulsations or the convection, or their coupling, for this relatively hot and luminous star. According to \citet{Houdek02}, for solar-type stars hotter than the Sun and with
masses $M\,>\,1.35\,M_{\sun}$, theoretical damping rates may be too small and estimated convective velocities too large, leading to predicted amplitudes larger than the data suggest.

We note however that the same scaling as in the solar case, $\nu_{\rm max}\sim0.6\,\nu_{\rm ac}$ where the acoustic cutoff frequency \mbox{$\nu_{\rm ac}\propto\,g\,T_{\rm eff}^{-1/2}$} \citep{Brown91}, can be used to obtain a good agreement for the frequency at which the oscillation amplitudes are largest. Using a recent value for the T$_{\rm eff} = 6530$\,K from \citet{Prieto02}, yields $\nu_{\rm ac}\sim\,1.7\,\mu$Hz and  
 $\nu_{\rm max}\sim\,956\,\mu$Hz.

\section{Conclusions}

Two-site observations of Procyon A confirmed the presence of solar-like oscillations, characterized by a large frequency separation 
$\Delta\nu_0=(53.6\,\pm\,0.5)\mu$Hz. We identified 45 \mbox{$p$-modes} frequencies in the range $300\,-\,1400\,\mu$Hz with harmonic degrees $l=0, 1, 2$, in good agreement with the theoretical predictions by \citet{Chaboyer99}.
The significative departure of the modes in the region of the maximum oscillation amplitudes from the asymptotic relation explains the difficulties in finding the mean principal frequency separation $\Delta\nu_0$ and the small frequency separation $\delta\nu_{02}$
 by usual techniques based on the best fit of the asymptotic relation from an $l$-mixed analysis.
Additional data with higher frequency resolution are needed to determine the rotational splitting and the
damping time of Procyon \mbox{\textit{p}-modes}. By determining oscillation frequencies with an accuracy better than $0.5\,\mu$Hz from an extended ground-based multi-site campaign on Procyon we will be able to test the techniques for the mode detection and prepare the scientific exploitation of future asteroseismic observations from space (MOST, COROT, EDDINGTON).

\begin{acknowledgements}
We thank Tim Brown for providing the AFOE pre-reduced data. We thank J. Schmitt for the ELODIE observations and collaboration.
We are grateful to OHP and Mt. Hopkins staff for the on site support of the observations. We are grateful to A. Baglin, J.L Bertaux, J.P. Sivan and A. Peacock  
for invaluable support for this project, P. Nisenson, R.W. Noyes, C. Barban and E. Michel for their contributions. We acknowledge INSU/PNPS for the financial support of the observations. MM was funded by ESA/SSD under contract C15541/01. 
\end{acknowledgements}

\bibliographystyle{aa}

\begin{thebibliography}{}
\bibitem[Audard et~al.(1995)]{Audard95}
Audard, N., Provost, J., \&  Christensen-Dalsgaard, J. 1995, A\&A, 297, 427

\bibitem[Baglin(2003)]{Baglin03}
Baglin, A. 2003, AdSpR, 31, 345

\bibitem[Baranne et~al.(1996)]{Baranne96}
Baranne, A., Queloz, D., Mayor, M., et al. 1996, A\&AS, 119, 373

\bibitem[Bedding \& Kjeldsen(2003)]{BeddingKjeldsen03}
Bedding, T.R., \& Kjeldsen, H. 2003, PASA, Vol. 20, p. 203-212

\bibitem[Barban et~al.(1999)]{Barban99}
Barban, C., Michel, E., Martic, M., et~al. 1999, A\&A, 350, 617

\bibitem[Bedding et~al.(2001)]{Bedding01}
Bedding, T.R., Butler, P.R., Kjeldsen, H.,  et al. 2001, ApJ, 549, L105

\bibitem[Bouchy et~al.(2001)]{Bouchy01}
Bouchy, F., Pepe, F., \& Queloz, D. 2001, A\&A, 374, 733

\bibitem[Bouchy \& Carrier(2001)]{BouchyCarrier01}
Bouchy, F., \& Carrier, F. 2001, A\&A, 374, L5

\bibitem[Bouchy \& Carrier(2002)]{BouchyCarrier02}
Bouchy, F., \& Carrier, F. 2002, A\&A, 390, 205

\bibitem[Brown et~al.(1997)]{Brown97}
Brown, T.M., Kennelly, E.J., Korzennik, S.G., et~al. 1997, ApJ, 475, 322

\bibitem[Brown et~al.(1994)]{Brown94}
Brown, T.M., Noyes, R.W., Nisenson, P., et~al. 1994, PASP, 106, 1285

\bibitem[Brown et~al.(1991)]{Brown91}
Brown, T.M., Gilliland, R.L, Noyes, R.W., et~al. 1991, ApJ, 368, 599

\bibitem[Carrier \& Bourban(2003)]{CarrierBourban03}
Carrier, F., \& Bourban, G. 2003, A\&A, 406, L23

\bibitem[Chaboyer et~al.(1999)]{Chaboyer99}
Chaboyer, B., Demarque, P., \&  Guenther, D.B. 1999, ApJ, 525L, 41

%
\bibitem[Christensen-Dalsgaard(2002)]{Christ02}
Christensen-Dalsgaard J. 2002, Rev. Mod. Phys., Vol. 74, 1073

\bibitem[Connes(1985)]{Connes85}
Connes, P. 1985, Ap\&SS, 110, 211

\bibitem[Connes et~al.(1996)]{Connes96}
Connes, P., Martic, M., \& Schmitt, J. 1996, Ap\&SS, 241, 61

\bibitem[Di Mauro \& Christensen-Dalsgaard(2001)]{dimauro01}
Di Mauro, M.P., \& Christensen-Dalsgaard, J. 2001, in Recent Insights into the Physics
of the Sun and Heliosphere: Highlights from SOHO and Other Space Missions, eds. P. Brekke, B. Fleck, \&
J.B Gurman, IAU Symp., Vol. 203, 94

\bibitem[Di Mauro et~al.(2003)]{dimauro03}
Di Mauro, M.P., Christensen-Dalsgaard, J., Kjeldsen, H., et~al. 2003, A\&A, 404, 341

\bibitem[Foster(1995)]{Foster95}
Foster, G. 1995, AJ, 109(4), 1889

\bibitem[Frandsen et~al.(2002)]{Frandsen02}
Frandsen, S., Carrier, F., Aerts, C., et~al. 2002, A\&A, 394, L5

\bibitem[Elsworth et~al.(1995)]{Elsworth95}
Elsworth, Y., Howe, R., Isaak, G.R., et~al. 1995, in GONG'94:
Helio and Astero-Seismology, eds. R.K Ulrich, E.J. Rhodes, Jr., \& W. Dappen,
ASP Conf. Ser., Vol. 76, 318

\bibitem[Gilliland et~al.(1993)]{Gilliland93}
Gilliland, R.L., Brown, T.M., Kjeldsen, H., et~al. 1993, AJ, 106, 2441

\bibitem[Girard et~al.(2000)]{Girard00}
Girard, T.M, Wu, H., Lee,et~al. 2000, AJ, 119, 2428

\bibitem[Grec et~al.(1983)]{Grec83}
Grec, G., Fossat, E., \& Pomerantz, A. 1983, Sol. Phys., 82, 55 

\bibitem[Guenther \& Demarque,(1993)]{GuentherDemarque93}
Guenther, D.B, \& Demarque, P. 1993, ApJ, 405, 298

\bibitem[Houdek et~al.(1999)]{Houdek99}
Houdek, G., Balmforth, N.J., Christensen-Dalsgaard, J., et~al. 1999, A\&A, 351,582

\bibitem[Houdek(2002)]{Houdek02}
Houdek, G. 2002, in Radial and Nonradial Pulsations as Probes of Stellar Physics,
eds. C. Aerts, T.R Bedding, \& J. Christensen-Dalsgaard, ASP Conf. Ser., Vol. 259, 447

\bibitem[Houdek \& Gough(2002)]{HoudekGough02}
Houdek, G., \& Gough, D.O. 2002, MNRAS, 336, L65

\bibitem[Kjeldsen et~al.(1995)]{Kjeldsen95}
Kjeldsen, H., Bedding, T., Viskum, M., et~al. 1995, AJ, 109(3), 1313

\bibitem[Kjeldsen \& Bedding(1995)]{KjeldsenBedding95}
Kjeldsen, H., \& Bedding, T. 1995, A\&A, 293, 87

\bibitem[Kjeldsen \& Bedding(2001)]{KjeldsenBedding01}
Kjeldsen, H., \& Bedding, T. 2001, in Helio- and
 Asteroseismology at the Dawn of the Millennium, eds. A. Wilson, P.L. Palle,
 ESA SP-464, 361

\bibitem[Kjeldsen et~al.(2003)]{Kjeldsen03}
Kjeldsen, H., Bedding, T., Baldry, I.K., et~al. 2003, AJ, 126(3), 1483

\bibitem[Lomb(1976)]{Lomb76}
Lomb, N.R. 1976, Ap\&SS, 39, 447

\bibitem[Martic et~al.(1999)]{Martic99}
Martic, M., Schmitt, J., Lebrun, J.C., et~al. 1999, A\&A, 351, 993 (Paper I)

\bibitem[Martic et~al.(2001a)]{Martic01a}
Martic, M., Schmitt, J., Lebrun, J.C., et~al. 2001a,
 in Recent Insights into the Physics of the Sun and Heliosp                                                                                                                                                                                                                            here -
Higlights from SOHO and Other Space Missions, eds. P. Brekke, B. Fleck, \&
J.B Gurman, IAU Symp. Vol. 203, 121

\bibitem[Martic et~al.(2001b)]{Martic01b}
Martic, M., Lebrun, J.C., Schmitt, J. et~al. 2001b, in Helio- and
 Asteroseismology at the Dawn of the Millennium, eds. A. Wilson, P.L. Palle,
 ESA SP-464, 431

\bibitem[ Matthews et~al.(2000)]{Matthews02}
Matthews, J., Kuschnig, R., Walker, G., et~al. 2002, in The Impact of
Large Scale Surveys on Pulsating Star Research, eds. L. Szabados, \&
D. Kurtz, ASP Conf. Ser., Vol. 203, 73 

\bibitem[Mozurkewich et~al.(1991)]{Mozurkewich91}
Mozurkewich, D., Johnston, K.J., Simon, R.S., et~al. 1991, AJ, 101, 2207

\bibitem[Nelder \& Mead(1965)]{NelderMead65} 
Nelder, J.A., \&  Mead, R. 1965, Computer Journal, Vol. 7, 308

\bibitem[Pijpers(2003)]{Pijpers03}
 Pijpers, F.P 2003, A\&A, 400, 241

\bibitem[Prieto et~al.(2002)]{Prieto02}
Prieto, C.A, Asplund, M., garci Lopez, R. J., et~al. 2002, ApJ, 567, 544

\bibitem[Provost et al.(2002)]{Provost02}
Provost, J., Martic, M., Berthomieu, G., et~al. 2002, in The
 First Eddington Workshop on Stellar Structure
 and Habitable Planet Finding, eds.
 B. Battrick, F. Favata, I. W. Roxburgh \& D. Galadi.,
 ESA SP-485, 309

\bibitem[Roberts et~al.(1987)]{roberts87}
Roberts, D.H., L\'{e}har, J., \& Dreher, J.W. 1987, AJ, 93(4), 968

\bibitem[Roxburgh \& Favata(2003)]{RoxburghFavata03}
Roxburgh, I.W.\& Favata, F. 2003, Ap\&SS, 284, 17

\bibitem[Scargle(1982)]{Scargle82}
Scargle, J.D. 1982, ApJ, 263, 835
%

\bibitem[Tassoul(1980)]{Tassoul80}
Tassoul, M. 1980, ApJ\&S, 43, 469

\bibitem[Walker et~al.(2003)]{Walker03}
Walker, G., Matthews, J., Kuschnig, R., et~al. 2003, PASP, 115, 000

\end{thebibliography}

\end{document}